**Rotational Multi-material 3D Printing of Soft Robotic Matter with Asymmetrical Embedded Pneumatics**

Jackson K. Wilt[1*], Natalie M. Larson[1,2*†], and Jennifer A. Lewis[1,3 †]

[1]John A. Paulson School of Engineering and Applied Sciences, Harvard University, Cambridge, MA 02138, USA.

[2]Department of Mechanical Engineering, Stanford University, Stanford, CA 94305, USA.

[3]Wyss Institute for Biologically Inspired Engineering, Harvard University, Cambridge, MA 02138, USA.

*Authors contributed equally to this work

† Corresponding authors

**Abstract:**

The rapid design and fabrication of soft robotic matter is of growing interest for shape morphing, actuation, and wearable devices. Here, we report a facile fabrication method for creating soft robotic materials with embedded pneumatics that exhibit programmable shape morphing behavior. Using rotational multi-material 3D printing, asymmetrical core-shell filaments composed of elastomeric shells and fugitive cores are patterned in 1D and 2D motifs. By precisely controlling the nozzle design, rotation rate, extrusion rate, and print path, one can control the local orientation, shape, and cross-sectional area of the patterned fugitive core along each printed filament. Once the elastomeric matrix is cured, the fugitive cores are removed, leaving behind embedded conduits that facilitate pneumatic actuation. Using a connected Fermat spirals pathing approach, one can automatically generate desired print paths required for more complex soft robots, such as hand-inspired grippers. Our integrated design and printing approach enables one to rapidly build soft robotic matter that exhibits myriad shape morphing transitions on demand.

**Introduction**

Soft robots possess many attributes that are difficult to replicate using conventional robots composed of hard materials. Most soft robots are produced by casting an elastomeric material onto a mold to pattern pneumatic channels on their surface, followed by lamination to a non-patterned elastomeric layer to encapsulate such channels. However, this multi-step approach to structuring soft materials involves tradeoffs between design complexity, fabrication time, and control.[1–3] Despite these tradeoffs, this approach has been widely used to build soft robots with fine-motor bending[1], multi-gait soft robots[2], high-strength pneumatics[3–5], and variable degrees of freedom actuators.[4] Each actuation characteristic was achieved using elaborate structuring of soft matter to elicit functionality, combined with conventional casting and laminating techniques.[6–8] This approach has recently been extended to 2D patterned surfaces fabricated using 3D printed molds, which result in mesostructured elastomers with more complex pneumatic channels.[9] These soft robotic materials exhibit complex out-of-plane curvature upon pneumatic actuation. Most recently, a novel fabrication route known as "bubble casting" was introduced.[10] This method relies on propagating a bubble through a tubular mold filled with an uncured resin, which results in an asymmetrical annular region due to differences between the density and viscosity of air (bubble) and the resin (liquid). Once patterning is completed, the elastomeric resin is cured, resulting in fluidically inflatable soft robotic fibers. While "bubble casting" is efficient and expands design possibilities, one must still produce molds for each design of interest.

Multi-material 3D printing is a rapidly growing approach for generating soft robotic matter that exhibits complex actuation modes.[11,12] To date, for example, millipede-like walkers [13] and fully autonomous soft robots[14] have been fabricated by direct and embedded printing, respectively. The ability to co-print elastomeric inks with tunable stiffness alongside fugitive inks that template fluidic conduits has opened new avenues for constructing soft robots. Recently, a new capability has emerged, known as rotational multi-material 3D printing (RM-3DP), which allows one to encode structural and functional properties at length scales below the characteristic filament diameter.[15] By creating intricate nozzles that permit simultaneous extrusion of multiple inks at varying rotation rates, one can generate soft materials with spatially tailored composition and properties with subvoxel control.[15,16] To date, RM-3DP has been used to fabricate elastomeric composites with programmed mechanics[15], dielectric elastomer actuators[15], and hydrogels with tunable swelling behavior. [17,18]

Here, we demonstrate the rapid design and fabrication of soft robotic matter with embedded pneumatics via RM-3DP. Using intricate nozzle designs coupled with controlled rotation rates, we co-print elastomeric and fugitive inks – the latter of which templates the pneumatic conduit designs of interest. The elastomeric ink is composed of a photopolymerizable polyurethane-acrylate resin, while the fugitive ink is a gel (30wt% Pluronic F-127 in water) under ambient conditions.[19] By controlling the location, size, and geometry of asymmetric pneumatic conduits, we can deterministically encode specific shape morphing transitions in both 1D filaments (e.g., 1D-to-2D and 1D-to-3D shape changes) and 2D objects (e.g., 2D-to-3D shape changes). We first generate 1D soft robotic filaments and characterize their mechanics through force and curvature measurements, which validate the computational modeling. Next, we showed that 1D and 2D soft robots can be fabricated that undergo shape-morphing transitions to 3D motifs. As a final demonstration, we harnessed a continuous filament print pathing algorithm that segments the structure into subdomains, using *connected Fermat spirals* [20] to generate print paths from images that directly encode preferential expansive strain directions. The integration of RM-3DP with automatic pathing both broadens the design space and allows one to achieve subvoxelated control of soft robotic materials.

**Results**

**Embedding asymmetrical pneumatics in soft robotic matter**

Simple elastomer fluidic actuators, such as the one shown in **(Fig. 1a)**, rely on the location and geometry of embedded asymmetric conduits to define their actuation direction and magnitude. Using RM-3DP **(Fig. 1b and Movie 1)**, we can program the conduit orientation, cross-sectional area, and geometry within printed elastomeric filamentary architectures. Nozzles composed of three distinct channels for ink flow **(Fig. S1)** allow one to print filaments that consist of a cylindrical elastomeric shell with internal elastomeric and fugitive ink-filled features **(Fig. 1b)**. The filament geometry is defined by the dimensionless flow rate $Q^* = \frac{Q}{v\pi R^2}$, where $R$ is the nozzle radius, $v$ is translational velocity, and $Q$ is the magnitude of ink flow rate. To customize the asymmetric conduit, the channel angle, $\varphi$, through which the fugitive ink flows **(Fig. 1c)** is modified by varying its angular offset from the nozzle bisecting line. While customized nozzles with either $\varphi=0°$ or $\varphi=45°$ are used for all experiments **(Fig. 1d)**, we carried out a detailed computational exploration of other $\varphi$ values and void shapes (**Fig. S2)**. To characterize the effects of programmed rotation, we also define a dimensionless rotation rate $\omega^* = \frac{R\omega}{v}$ which captures the helical character of continual reorientation using nozzle radius, translational velocity, and rotational velocity ω.[15] The nozzle rotation rate can be varied between zero (ω*= 0, no rotation **(Fig. 1e)**), discrete steps, or continuous rotation (ω*= -1, **(Fig. 1f)**) during printing. Key parameters defining the sub-voxelated filament architecture are the dimensionless flow rate, $Q_f^*$ of the fugitive ink and the orientation of the templated conduit relative to the filament major axis and substrate, $\theta_f$ **(Fig. 1g-i)**. The orientation component is set as $\theta_f$ with directional adjustments assigned as clockwise ($-\Delta\theta_f$) or counterclockwise ($+\Delta\theta_f$) based on programmed radial translations.

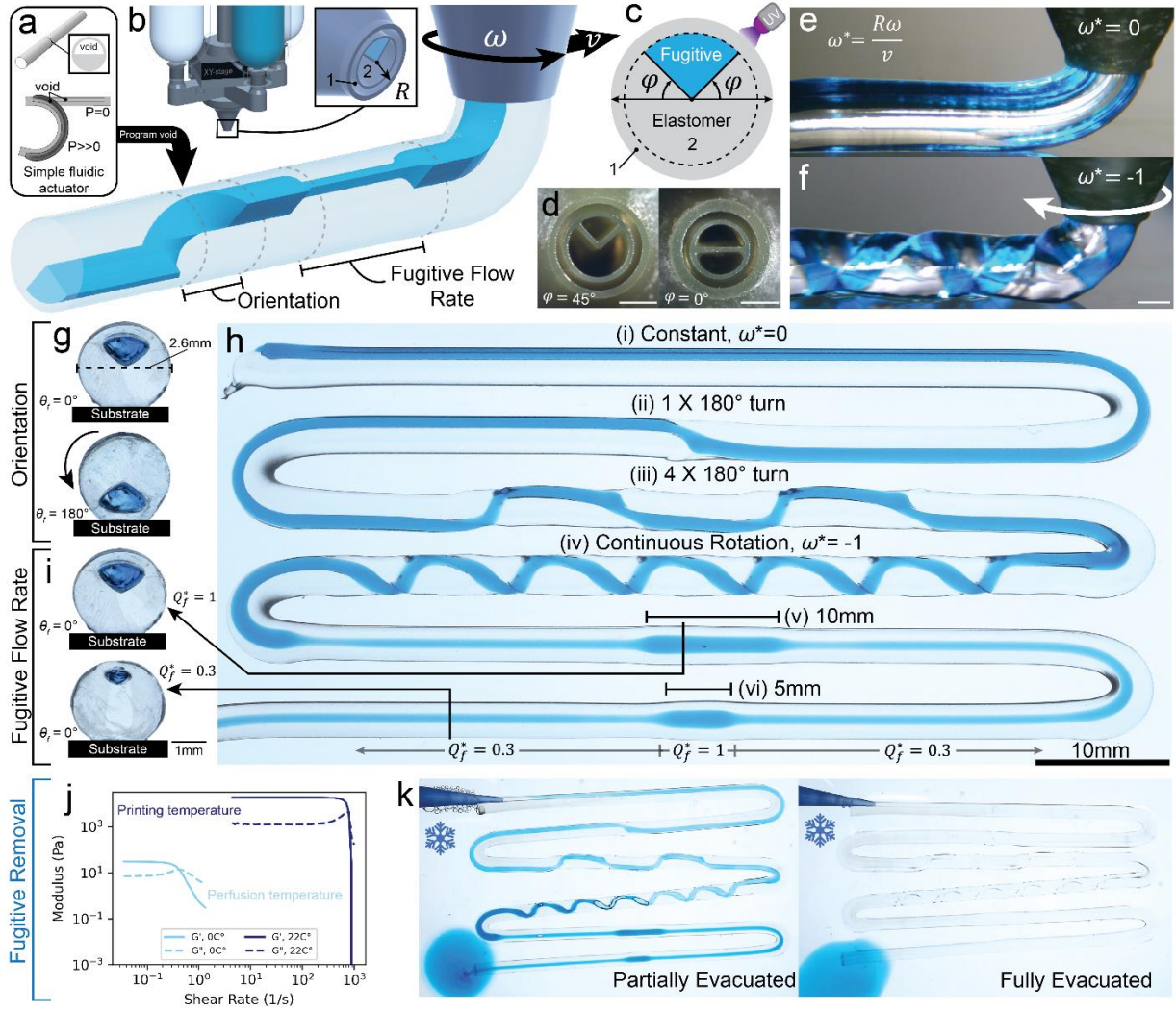

**Figure 1**: **Rotational multimaterial 3D printing of soft robotic filaments with embedded pneumatics.** (a) Schematic view of a simple fluidic actuator showing the effect of an asymmetric pneumatic conduit on bending. (b) Rendered image of rotational 3D printing printhead that enables co-printing of elastomeric and fugitive inks with customized nozzles at controlled ink flow rates. (c) Illustration of the printing nozzle outlet showing the fugitive ink-filled sector defined by $\varphi$ and (d) optical image of the nozzle outlet with pie-shaped $\varphi=45°$ (left) and hemispherical sectors $\varphi=0°$ (right). Side profile of filament printing at a dimensionless rotation rate of (e) $\omega^*=0$ and (f) $\omega^*=-1$. (g) Cross-section images of printed filaments with $\Delta\theta_f = 0$ (top) and $\Delta\theta_f = +180°$ (bottom) configurations. (h) Continuous filament printed in a serpentine pattern with varying $\omega^*$ and $\Delta\theta_f$. (i) Effect of normalized fugitive flow rate $Q_f^* = 1$ (top) and $Q_f^* = 0.3$ (bottom) on the cross-sectional area of the fugitive ink-filled conduit. (j) Storage (G') and loss (G'') moduli as a function of shear rate under ambient (printing) and cooled (fugitive ink removal) temperatures. (k) Optical images of fugitive ink removal from the printed filaments, in which an aqueous 30wt% Pluronic ink is removed by pumping cold water through the filament. (Scale bar = 1mm unless otherwise denoted). [Note: All printed filaments have a diameter of D=2·R=2.6mm ± 0.06mm and are printed at a velocity v=3mm/s and print height of h=3.1mm where h=2·R and h*=h/(2·R) =1.2)].

By systematically varying the fugitive ink flow, $Q_f^*$ and printhead orientation, $\theta_f$, we can print continuous filaments with asymmetric pneumatic conduits whose cross-sectional area and orientation vary along the filament major axis **(Fig. 1h)**. Initially, the filament is printed at a constant flow cross-section $Q_f^* = 1$ with discrete and continuous programmed reorientations **(Fig. 1h (i-iv))**. We show a printed filament with a void located at the "top" of the filament ($\theta_f = 0°$) **(Fig. 1g (top)** and **Fig. 1h(i))**. Next, the printhead orientation is changed to $\Delta\theta_f = +180°$ to locate the void at the "bottom" of the filament **(Fig. 1g, bottom)**. To incorporate multiple reorientations $\Delta\theta_f$ with distinct regions, the printhead must be rotated over a prescribed transition length, $L_t$. A printed filament segment with one such transition $\Delta\theta_f = +180°$ is shown in **Fig. 1h(ii)** for $L_t$=2.5mm, which is comparable to the nozzle diameter. Another filament segment with four $\Delta\theta_f = +180°$ reorientations is shown in **Fig. 1h(iii)**, resulting in sequentially patterned regions of void location. One can also impose a continuous rotation ($\omega^* \neq 0$), see for example, the filament segment produced at $\omega^*= -1$ in **Fig. 1f** and **Fig. 1h(iv)**. This parameter characterizes a continuous subvoxel property assignment with a nominally helical motif. The cross-sectional area of the asymmetric pneumatic channel can be modulated by varying the printing pressure applied to the fugitive ink **(Fig. 1i)**. Using a reduced normalized flow rate $Q_f^*<1$ results in a smaller fugitive ink-filled void, while still retaining angular location of the channel at a flow rate of $Q_f^*=1$. For fugitive flow rates of $Q_f^*=1$ **(Fig. 1i, top)**, the nozzle outlet and fugitive ink-filled channel dimensions are identical, which corresponds to the maximum fugitive flow rate and applied pressure explored. Using a combination of $Q_f^*=1$ and $Q_f^*=0.3$ (85% maximum syringe extrusion pressure) over two different lengths results in the last two filament segments shown in **Fig. 1h(v-vi)**. This combination results in a fugitive ink-filled conduit in the $Q_f^*=0.3$ region that rapidly transitions (~$L_t$ at v=3mm/s) from $Q_f^*=1$ back to $Q_f^*=0.3$ along these segments. To complete the fabrication process, the elastomeric ink is first crosslinked by UV light followed by removing the fugitive ink by cooling the printed architectures to 0°C **(Fig. 1j)**. Under these conditions, the storage modulus ($G'$=30Pa) of the fugitive ink is significantly reduced **(Fig. 1k)**. One can fully remove the fugitive ink leaving behind a continuous asymmetric pneumatic conduit of programmed shape and orientation upon submersion in ice water and subsequent perfusion **(Movie 2)**.

**Programmable shape morphing and mechanics of 1D filaments**

We first demonstrate the actuator response of a single straight filament with a constant cross-section flow rate $Q_f^*=1$ and nozzle outlet angle $\varphi = 45°$ **(Fig 2)** and to a lesser extent $\varphi = 0°$ **(Fig. S3)**. We pressurize the filaments and perform curvature measurements, to extract $\kappa=1/r$ along the actuator length **(Fig. 2b)**. For comparison, we also simulate actuation using the built-in hyperelastic fitting module, which aligns with an Ogden hyperelastic model (Abaqus CAE, Dassault Systèmes) **(Fig. 2c)**. Computational and experimental results are in good agreement, albeit with slight underestimation of curvature at the highest pressures up to 103kPa tested **(Fig. 2d)**. We measured the thickness of the shell adjacent to the void as a function of the central angle, $\theta_{shell}$, within the region of the void **(Fig. 2e)**. The shell is thickest near the edges of the void (0.36 mm), and thinnest in the center (0.30 mm). We also carried out actuation cycles to determine the average curvature along the filament length as a function of increasing pressure **(Fig. 2f)**. Notably, the printed filaments did not fully recover their original shape upon actuation at 97kPa, and they exhibit rapid ballooning and rupture above 103kPa.

Next, we characterized their force generation and response to the loading. The force output data show that patterning single actuator lines adjacent to one another resulted in increased total force **(Fig. 2g)**. For a single filament actuator (25mm in length), force generated is 10.3mN at 10.3kPa. Above this applied pressure, the filaments exhibit out-of-plane bending and permanent strain deformation. We then patterned parallelized actuators to create a more stable deformation profile. Architectures with two and four parallel filaments exhibited higher respective values of output force, 31mN and 52mN, at 103kPa. For 25mm filaments lifting 2g weights, a non-uniform, coiling actuation is observed **(Fig. 2h)** and the curvature calculated by tracking the filament centerline and approximating average curvature across the filament length. By contrast, shorter filaments (7.5mm in length) exhibit greater curvature, which remains in-plane **(Fig. 2i)** but is accompanied by more significant shell expansion. To "lock in" their displacement upon actuation, we infused the embedded pneumatic conduit with a photocurable resin that becomes rigid upon UV curing. When transformed in this manner, a 25mm filament can hold a 200g load **(Fig. S4 and Movie 3)**. The skeletal-based motifs open new avenues to creating hybrid (stiff-soft) robotic systems.

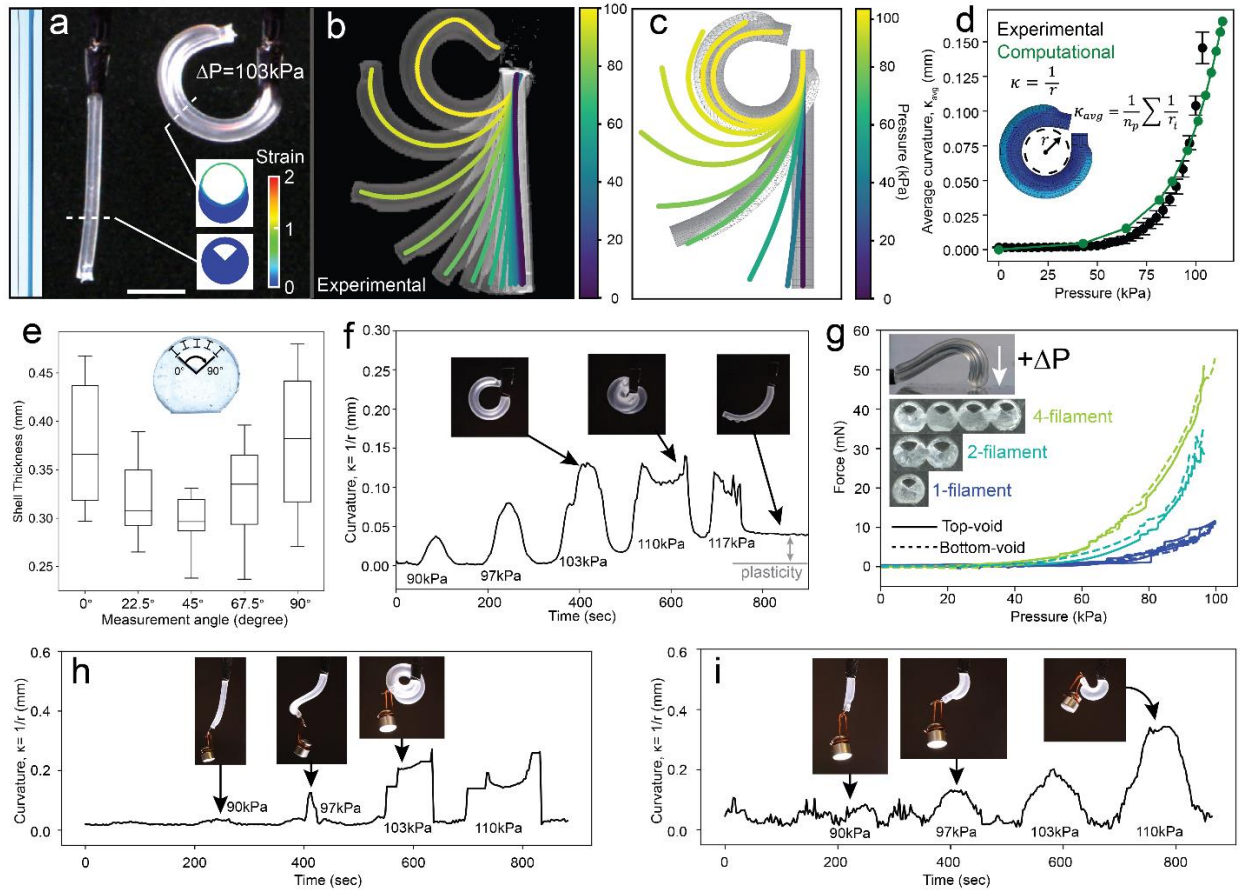

**Figure 2: 1D filaments with embedded asymmetric pneumatics**. (a) A single, $\varphi = 45°$, 25-mm filament is actuated with an accompanying simulation showing internal undeformed and deformed cross-sections. We incrementally increased the pressure in the filaments by 69kPa with a pressure ramp of 1.7kPa per second (b) The experimental series of tracked filament midpoints and exported overall averaged curvature from all filamentary points. (c) Computationally derived midpoint values of a 2.6mm diameter filament and corresponding curvature colormap. (d) Comparison between the experimentally determined and computational curvature maps under gravity to validate model-based curvature output (n=3). (e) Filament-shell thickness distribution gathered from a perfused cross-section (n=9). (f) Curvature tracking of free-hanging 25mm filament actuated at increasing pressures with regions of plastically retained curvature. (g) Coupled filaments force amplification using single, double, and quadruple parallelized bending actuators. (h-i) Lifting and work performance of increasing pressure-induced actuation ($\varphi = 45°$) with a 2g weight using two filament lengths: (h) 25mm and (i) 7.5mm (n=1). (scale bar is 10mm)

We actuate printed filaments with programmed variations in void orientation and cross-sectional area **(Fig. 3)** to demonstrate controlled tuning of actuation direction and magnitude along 1D filaments. First, we programmatically controlled actuation direction using discrete **(Fig. 3a-b)** and continuous **(Fig. 3c)** changes in void orientation ($\theta_f$) **(Movie 4)**. Using volume-controlled actuation with a pneumatic syringe, we examined a filament with a $\Delta\theta_f = +180°$ twist at its midpoint **(Fig. 3a)**. Upon actuation, this filament produces two curvatures $+\kappa$ and $-\kappa$ equal in magnitude but in opposite directions within the same plane. Building on this capability, we then implemented four discrete $\Delta\theta_f = +180°$ rotations with 10mm bending sections to produce soft robotic filaments with periodic changes in curvature direction ($+/-\kappa$) upon actuation **(Fig. 3b)**. As a final example, we created a filament with continuous changes in orientation corresponding to $\omega^* = -1$ in **(Fig. 3c)**. Upon actuation, this filament bends and twists into a coil shape, demonstrating 880° of angular displacement at the free end at 83kPa. Computational modeling is in good agreement with the experimentally observed deformations using identical geometry and previously derived Ogden mechanical models **(Fig. 3a-c)**.

The fugitive flow rate $Q_f^*$ is used to explore how discrete variations in the area of the pneumatic conduit impacts local curvature. By varying $Q_f^*$, we created curvature changes along a single programmed filament. For instance, a hinge actuator is fabricated with an initial 25mm segment at $Q_f^*=0.3$, proceeded by a 10mm region at $Q_f^*=1$, and followed by another 25mm segment at $Q_f^*=0.3$. Once cured and actuated, this configuration produces distinct bending behavior **(Fig. 3d)** with the filament going from a hinge angle $\theta_h = \sim180°$ at 0kPa and $\Delta\theta_h = +180°$ at 103kPa. Next, we compared two hinged actuators with 25mm "inert" regions on both sides of the hinge regions, exploring both 5mm and 10mm hinge lengths printed at $Q_f^*=1$. During pressurization, the $Q_f^*=0.3$ regions act as inert regions, while the $Q_f^*=1$ regions demonstrate greater bending. To quantify this behavior, the 5mm and 10mm hinge actuators are pressurized in a quasistatic process, enabling precise measurement of the hinge angles **(Fig. 3e)**. The 10 mm hinge actuates at lower pressures, beginning to bend at ~34kPa and then proceeded in a relatively linear fashion in terms of plotted angle up to $\theta_h = \sim0°$ at 103kPa. By contrast, the 5 mm hinge remains largely inactive until 83kPa. While the 5mm hinges required higher pressures, they exhibit sharper bends over a shorter pressurization range. Both hinge lengths reach a maximum curvature of $\kappa=0.5$ **(Fig. 3f)**, but the 5mm hinges experience rapid inflation in the 97–103 kPa range, achieving a maximum hinge angle of $\theta_h$= ~90°.

By programming void orientation and fugitive flow rate, it is possible to fabricate a more complex 3D geometry, i.e., a wireframe cube, within a single-filament soft structure **(Fig. 3g)**. Using finite element modeling, we design a rod with six consecutive hinges, incorporating three pairs of adjacent high-flow regions oriented to actuate within the same bending plane. The hinge structure consists of inert 25 mm struts and 5 mm hinge regions, each incorporating two $\Delta\theta_f$=+90° after every two hinge locations. We then printed an analogous actuating wireframe structure and performed volume-controlled actuation using a pneumatic syringe for more precise control over inflation and to avoid hyper-expansion during three actuation states: unactuated **(Fig. 3h, top)**, partially actuated **(Fig. 3h, middle)**, and fully actuated **(Fig. 3h, bottom and Movie 5)** up to a maximum pressure of 83kPa. Notably, the final 3D actuated state exhibits slight deviations in uniform hinge angles, primarily due to self-weight interactions with the substrate. Higher-elastic materials may enable more complex deformations and far-field actuation effects in a 1D soft rod because of their resistance to self-weight distortion effects.

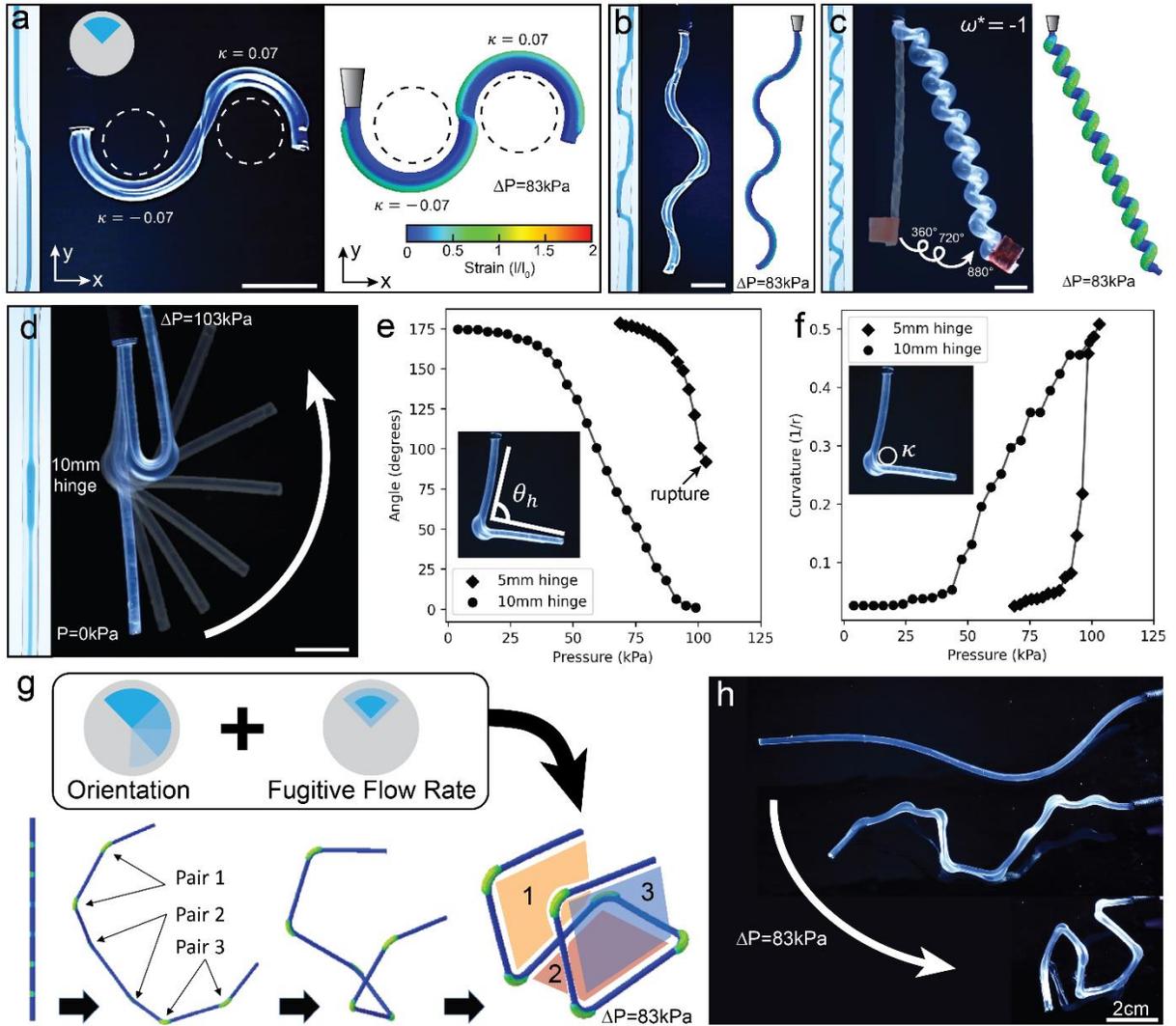

**Figure 3**: **Orientation and fugitive ink flow rate effects on complex filament actuation.** (a) Computational and experimental filament actuation with a single $\Delta\theta_f$=+180° showing two bending modes, (b) four $\Delta\theta_f$=+180° reorientations showing five bending modes, and (c) continuous $\Delta\theta_f$ change using $\omega^*$= –1 with coiling behavior. (d) A pneumatic 10mm hinge structure (e) angle and (f) curvature, $\kappa$=1/$r$, with respect to pressure for both a 5mm and 10mm hinge. (g) Orientation and fugitive extrusion rate processes combined to produce three pairs of hinges with independent bending planes, resulting in a simulated cubic wireframe structure. (h) Experimental volumetric inflation of the hinged filament demonstrating the deformation to an actuated cubic wireframe configuration. (unless otherwise denoted, the scale bar is 10mm).

**Programmable shape morphing of 2D architectures**

Our RM-3DP method also allows filaments to be printed in 2D architectures. When inter-filament contacts are minimal, e.g., contact only at crossover points, we observe bending along the filament axis akin to our 1D actuators **(Fig. 4 a-b)**. As an initial example, consider the structure in **Fig. 4a** which consists of a two-layer logpile with voids positioned all along the top ($\theta_f$=0º) for the top layer and bottom ($\theta_f$=180º) for the bottom layer, $Q_f^*$=1, and interfilament spacing of 200% of the filament diameter to ensure unconstrained inflation. Each layer is fluidically independent and actuated individually to produce two distinct bending surface directions. We printed this 2D structure and showed that its actuation behavior is in good agreement with simulations **(Fig. 4b, Movie 6)** at a pressure up to 103kPa.

For 2D architectures composed of many adjacent filaments, lateral expansion of filaments with $\varphi = 0º$ (half-circle) causes significant bending perpendicular to the filament axis **(Fig. 4c)** as shown using computational modeling. By changing the pneumatic conduit geometry to $\varphi = 45º$ (quarter-circle), lateral expansion is reduced **(Fig. 4d)** and the predominant bending direction is parallel to the filament axis. We observe opposing moments in reoriented surface actuators when combining nozzle rotation and inter-filament contact. To compare the effects from differing orientation regions, each printed surface consists of square cells containing alternating filament orientation with distinct bending properties, initially validated using Abaqus simulation **(Fig. 4e-f)**. Each cell consists of ten printed parallel lines with a top void, $\theta_f=0°$, each 25 mm in length. After the first ten lines, a reorientation of $\Delta\theta_f=+180°$ occurs, transitioning the void parallel (∥ transition) to the print direction and deposits 10 lines, $\theta_f=180°$, each 25 mm in length. When actuated, pneumatic conduits that are parallel to each other, but with opposite orientation, constrain deformation at the cell interface **(Fig. 4e)**. By contrast, when a $\Delta\theta_f=+180°$ reorientation perpendicular (⊥ transition) to the print direction is utilized in bending along the filament axis is unconstrained **(Fig. 4d)**.

By coupling both reorientation modes, we show a printed 3x3 checkerboard shape with alternating regions of intended $\theta_f=0°$ and $\theta_f=180°$ sequences **(Fig. 4g)**. Each printed cell contains regional $\theta_f$ values that are then tested for inflated bending mode directions. Using the same cellular assignment approach we created multiple complex muti-modal actuating surfaces in addition to the exemplar 3x3 surface in **Movie 7**. One surface produced included an internal 10 filament-wide square region with positive bending $\theta_f=180°$, surrounded by a 5 filament-wide border with a negative bending direction, with $\theta_f=0°$, shown in **(Fig. 4f, left)**. We also fabricate a 2x2 **(Fig. 4f, middle)** and 3x3 **(Fig. 4f, right)** checkerboard pattern using alternating $\theta_f=180°$ and $\theta_f=0°$ regions that contain a distribution of unconstrained and constrained bending. These results reveal the effectiveness of programmatic control over regional bending, while simultaneously underscoring the challenges associated with competing bending interactions in multi-modal actuating surfaces. Future studies will explore optimized transition strategies to enhance functional deformation control.

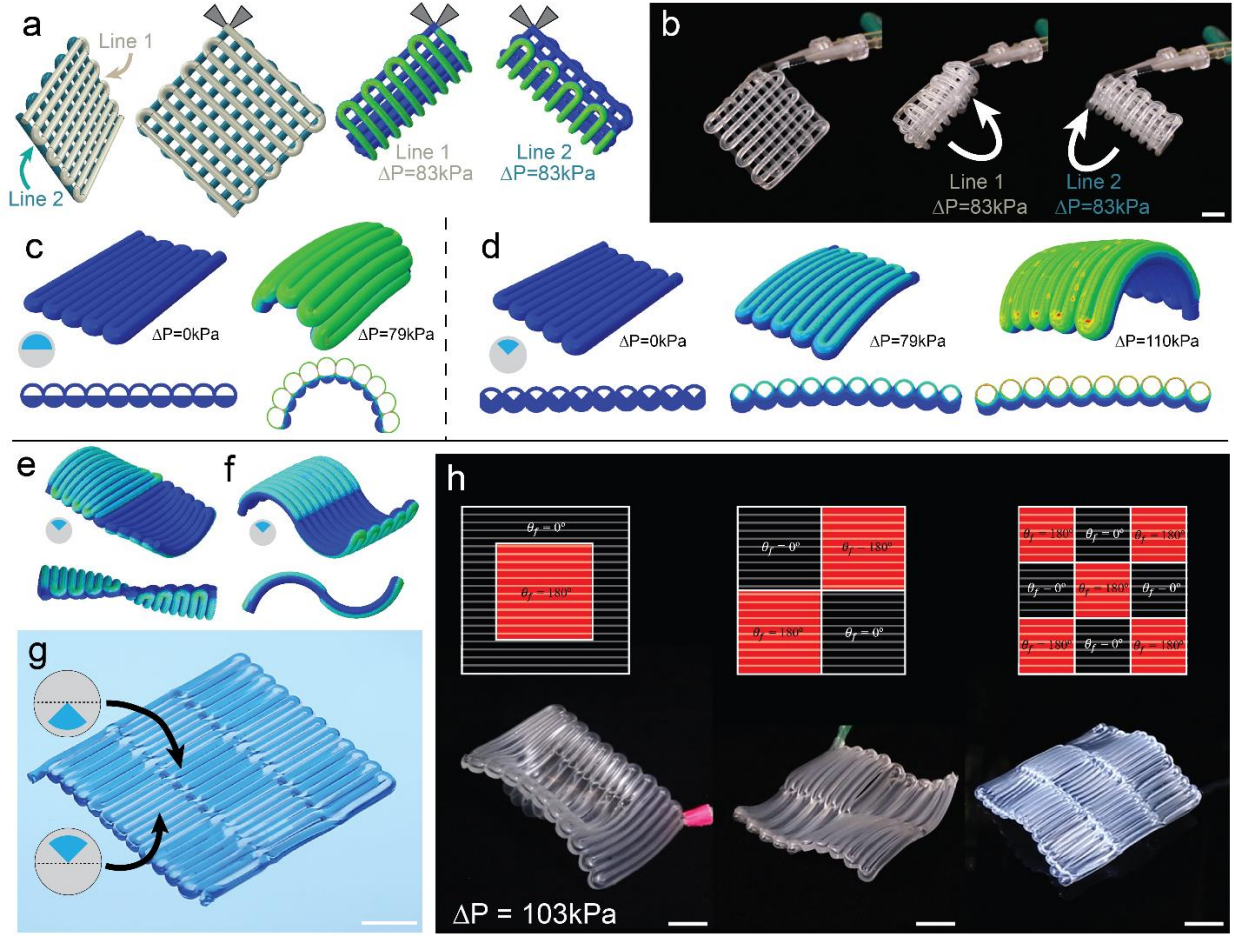

**Figure 4: Shape morphing behavior during actuation of 2D architectures**. (a) Simulated dual-input bi-layer structure with opposing directions and orthogonal print lines showing opposite bending directions. (b) Dual-input bilayer structure with each of the single lines pressurized and achieving two final actuated states. (c) Simulated semi-circular cross section ($\varphi=0°$) parallelized filament surface. (d) Simulated quarter-circle cross-section ($\varphi=45°$) parallelized filament surface at increasing pressurizations. (e) Parallelized actuating lines with the inversion of void location parallel to the primary void direction. (f) Parallelized actuating lines with the inversion of the void location perpendicular to the primary void direction. (g) Checkerboard patterned 3x3 with alternating $\theta_f=0°$ and $\theta_f=180°$ void location and thus bending direction of cells. (h, left) Actuating surface with a $\theta_f=180°$ bending center surrounded by $\theta_f=0°$ direction. (h, middle) Empirical integration of both non-interfering and frustrating modes of the surface of a 2x2 actuating checkerboard. (h, right) Checkerboard patterned 3x3 with alternating positive and negative bending cells. (scale bar is 10mm)

**Algorithmic continuous print pathing for building soft robotic matter**

Continuously pathing arbitrary structures is a fundamental goal of 3D printing, particularly for fabricating complex biological or personalized shapes. Zhao et al. initially proposed an algorithm for connected Fermat spiral pathing [20] to improve efficiency in FDM printing of complex structures. Building on this approach, we adapted the continuous Fermat pathing toolkit to generate single-layer continuous fluidic pathways. Pathing and orientation are dictated entirely by the input geometry, as the Fermat pathing algorithm generates isocontours and isolated regions, which are then continuously interconnected **(Movie 8)**. Previously, we defined $\theta_f$ as an absolute coordinate system. We now introduce $\psi$ to denote relative incremental angle changes that modify the initialized $\theta_f$ configuration.

We use Python scripting to produce the print pathing coupled with the orientation of the printhead that encodes the direction of bending during continuous deposition. This workflow is demonstrated using a vectorized image of a six-petaled flower **(Fig. 5a)** accompanied by an angle colormap that maintains the fugitive material in the top region ($\theta_f=0°$), while incrementally adjusting $\psi$ based on the direction of the printhead as it is translating. Each angle is stored and procedurally applied to the reorientation G-code along the printhead rotational axis in relative coordinates. The flower structure is printed, cured, perfused, and fully sealed. A dispensing tip is inserted into the center filamentary ring to connect a pneumatic line to the continuous fluidic structure. The flower structure **(Fig. 5b, Movie 9)** is pneumatically pressurized using intermediary pressure (55kPa) for partial actuation and higher pressure (69kPa) to elicit pronounced actuated petals. The flower is used as a simple example of our ability to use image-based methods to directly build soft robotic matter.

As a further demonstration, we imported an RGB image of a supinated human hand **(Fig. 5c)** which is transformed via vector graphics software (Adobe Illustrator), where each of the digits of the hand is replicated and extended to the base of the wrist using vectorized polygons **(Fig. 5d)**. Once the individual digits are imported, we incorporate the orientation protocol **(Fig. 5e)**, and program a top void ($\theta_f=0°$) for all connected Fermat spirals according to the $\psi$ correlation colormap. The hinge locations are selected based on the knuckle joint locations on the supinated hand and defined by area-based selection around the joint. We then print the rotationally parameterized filaments and hinges to produce a surface-actuating robotic hand **(Fig. 5f)**. The hand is postprocessed and then flipped 180° onto its other side to observe the upward bending of the digits. The soft robotic hand digits are subsequently controlled **(Fig. 5g and Movie 10)** using pressure regulators at 83kPa to elicit the significant angle response for each finger. The actuating hand is shown counting in the image sequence index, middle, ring, and pinky fingers. This motif incorporates all programmatic aspects of RM-3DP, including reorientation, fugitive flow rate, and adaptive continuous pathing.

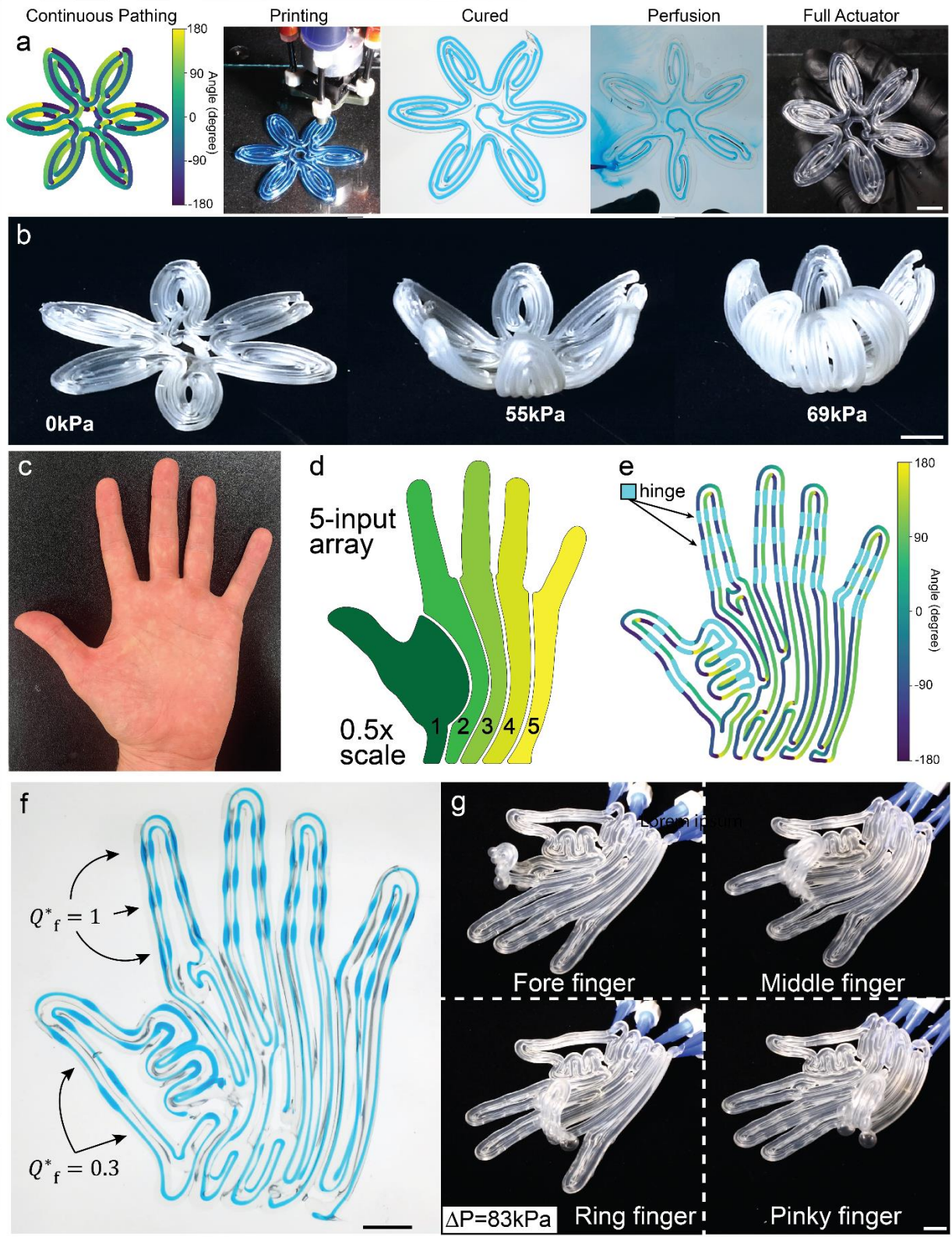

**Figure 5: Image-based print-path planning for generating complex soft robotic matter.** (a) Continuous pathing workthrough for fabricating a continuously pathed soft robotic 6-petaled flower. (b) Actuation sequence of printed

flower structure from 0kPa to 69kPa. (c) Image of a supinated human hand. (d) Partitioned hand with a color gradient indicating distinct digit regions from the image of a supinated hand. (e) 5-input array processed using the continuous Fermat pathing tool, with the associated direction of printing encompassed by the printing angle and regions with a regular extrusion rate of the fugitive. (f) The fully printed hand actuator before fugitive removal. (g) 3D printed multi-input hand and actuated four separate independent channels that actuate in sequence index, middle, ring, and pinky finger. (scale bar is 10mm)

## Summary

We have shown that rotational multimaterial 3D printing can be used to rapidly fabricate filamentary architectures with asymmetric pneumatic channels that are capable of 1D-to-3D shape transformations. To expand beyond rectilinear pathing, we used connected Fermat spirals to path organic contours and regionally assign more complex 2D configurations, including a vectorized "flower" image and a supinated hand containing hinge regions for recapitulating biomimetic movement. Using this novel 3D printing method that integrates multi-material printheads with programmed rotation, one can create soft robotics and other deployable structures for myriad applications.

## METHODS

**Materials**

The elastomeric acrylic ink material is prepared by mixing 37.14g of urethane oligomer (CN9018, Sartomer), 37.14g of isodecyl acrylate (Sigma Aldrich), 0.742g of the 2,2-dimethoxy-2-phenylacetophenone (Irgacure 651) photoinitiator, and 14.84g of fumed silica nanoparticles (CAB-O-SIL TS-720, Cabot Corp.), which acts as a rheological modifier. Mixing uses a planetary centrifugal shear mixer or SpeedMixer (FlackTek, Inc.), with the materials placed in disposable polypropylene cups. All material components and 7.42g of the fumed silica (half of the total fumed silica mass) are added together and undergo two instances of 2000 rpm mixing for 2 min with cooling for 2 min between the mixing stages. After mixing half of the total fumed silica, the ink is manually reintegrated into the base of the mixing cup. Then, an additional 7.42g of fumed silica is added (total fumed silica mass) for another full mixing cycle. After all the fumed silica has been added and two mixing cycles have been performed, the mixture is manually reintegrated for a third mixing stage to ensure all solid agglomerates have been dispersed. The elastomeric acrylic inks are manually loaded using a spatula into 55-cc UV/light block amber syringes (Nordson, EFD). The loaded syringe is centrifuged (Avanti centrifuge J-25 I, Beckman Coulter Avanti Inc.) at 2500rpm for 10 min to remove trapped air.

The fugitive ink is a perfusable poloxamer gel consisting of 30wt% Pluronic F-127 dissolved in deionized water. The solid Pluronic F-127 powder at 86g is placed in 200ml of deionized water along with 0.4g of liquid blue food coloring and manually shaken briefly to enable the powder to be moderately coated in water. The solid-liquid mixture was placed in a laboratory refrigerator (at or below 5°C) for 48 hours. If agglomerates were still present, the solution was shaken and left in the refrigerator for an additional 24 hours. The rheological profile **(Fig. 1j)** of the 30w.t.% Pluronic fugitive ink, at a printing temperature of 22°C, has a robust storage modulus plateau at 1,900Pa with a favorable shear thinning and yield stress response while printing. In concert with the favorable elastomeric viscoelastic and yield stress ink properties, the fugitive can be deposited and retain its geometric accuracy in the filament's uncrosslinked form. In its cool and more liquidous phase, we loaded 55-cc clear syringes by pouring the material directly from the mixing cup (Nordson, EFD). No centrifuging was required because the liquidous phase did not contain trapped air. The filled syringe is left to reach room temperature for a minimum of several hours. For filaments produced in this work (5-100cm in continuous length), typically 3-5 perfusions of 5cc syringes of ice water were sufficient to remove the fugitive ink

**Multi-material printheads**

The asymmetrical core-shell nozzle was designed in a 3D computer-aided design software (Autodesk Fusion 360). The nozzles were exported as stereolithography (STL) files and printed on a commercial direct light projection printer (D4K printer, EnvisionTEC) using HTM140 resin (EnvisionTEC). The geometry of the nozzle was precisely selected based on the dimensional XY-resolution of the printer, which is approximately 25μm. The outer diameter is 2.8mm with a shell thickness of 300μm **(Fig. S1)**. The printheads are reused for months at a time because the urethane oligomer ink is stable for extended periods, and each time a new ink formulation is added, the heads are fully perfused with the freshly mixed ink.

**Rotational 3D Printing**

The printing speed, extrusion rate, and printing height determine the final desired filament geometry because the ink relies on its viscoelastic and yield stress properties. We avoid deforming the filaments on the substrate by allowing the structure to be fabricated at a reduced cross-sectional area compared to the constant flow rate $Q^*=1$. The print velocity is $v=3$mm/s for volumetric flow rate $Q^*$ and $h^*$, and the print height is $h^*=1.2$ of the nozzle outer diameter. Initial tests showed that increased speed with respect to constant flow rate improves $Q^*=1$ filament geometry retention by avoiding substrate interaction. This effect is mainly due to the plastic deformation of the filament as it is being deposited onto the substrate, as the material's yield stress is adequate for a self-supporting filament but induces geometric deviation at smaller heights. This produces a final filament geometry of 2.6mm and a shell thickness of 300-360μm **(Fig. 2e)**. Our surface printing process decreases fugitive extrusion $Q_f^*=0.3$ during acute angle turns, approximately $\Delta\theta_f = +180°$ over 2mm, and prevent small shell thickness that can lead to premature ballooning.

Once the ink is printed, the structural acrylic is then fully cured under a UV-light lamp (Omnicure Series 2000, 200W power and 320-500nm wavelength, Lumen Dynamics Inc.) for a minimum duration of 5min under an ambient Argon (Airgas Inc.) atmosphere for full cross-linking. The post-cure processing involves submerging the photo-cured structures in ice water (~0°C) for 30 seconds **(Fig. 1k)**. The gel transitions to a more viscous phase and is perfused using the same ice water from a 10cc syringe, which evacuates the viscous fugitive material out the other end of the channel. Once the fugitive is perfused, we use airflow to evacuate any liquid left in the channel. The device is subsequently sealed at the end of the pneumatic line with additional photocuring acrylic elastomer (backfilled by 2.5mm into the network with a handheld ink syringe), which is then photocured using UV-light for another 300 seconds under an argon atmosphere. The final robotic structure is outfitted with dispensing tips (Nordson, EFD) connected to a pressure regulator line for pressure-based control or a pneumatic syringe for pneumatic control.

**Actuation of 1D and 2D filamentary architectures with asymmetric pneumatics**

We use a length-to-diameter (*L:D*) ratio of approximately 10:1, resulting in 25mm long filaments. The length of 25mm is also selected to avoid substantial self-weight effects, whereas, with longer filaments ($L \gg 25$mm), gravity induces out-of-plane contraction and nonlinear distortion. The initial filament actuators are measured under a 0.35 kPa/sec ramping pressure to emulate quasistatic pressurization using pneumatic pressure regulators (PCD-100PSIG-D, Alicat Scientific). We tested the force generated from the filaments in a horizontal configuration to demonstrate an expansion and pressing of the baseplate. The baseplate is coated in silicone oil to evaluate the relative amount of vertical force exertion and allow the actuator to slide according to horizontal forces. All other curvature actuators are tested free-hanging in open air whereas all surface actuators are tested on a lubricated baseplate of acrylic coated in silicone oil. We use volumetric inflation in conjunction with pneumatic pressure-induced actuation where the air pressure is not denoted. Volumetric air inflation enables more precise control over expansion and nonlinear ballooning failure that occurs in some samples, which are denoted in their relevant sections. Actuation using volumetric air injection is conducted with 10mL pneumatic syringes, which provide a substantially larger reservoir than the filament void. Applying Boyle's Law ($P_1V_1 = P_2V_2$), we calculate the applied pressure based on plunger displacement.

**Simulation**

The simulation was conducted in Abaqus using a fitting of an Ogden hyperelastic model:

$$W = \sum_{i=1}^{N} \frac{\mu_i}{\alpha_i}(\lambda_1^{\alpha_i} + \lambda_2^{\alpha_i} + \lambda_3^{\alpha_i} - 3) + \frac{1}{D_i}(J-1)^2$$

Where $\mu_i$ and $\alpha_i$ are the material parameters. $N$ represent the number in the specified model and in this instance $N=2$. The $D_i$ is the bulk modulus. The fitting is conducted on the uniaxial testing data from Larson et al. and using the same soft acrylic ink.[15] The output model parameters are $\mu_1$ =0.0858, $\mu_2$ =0.0467, $\alpha_1$ =2.73, $\alpha_2$ =-0.73, $D_1$=0 and $D_2$=0.

**Continuous Print Pathing**

The Python codebase for the connected fermat spirals pathing can be accessed in terms of the analysis work at [google collab link](). The Python codebase uses Plotly to assign the properties of the interpolated G-code function. The Python codebase (see Continuous Pathing methods section), where the individual digits in the connected Fermat Path form are imported as XY-collated points and are scaled to half of the photographed hand size and placed at 90% one-filament diameter separation between each digit to ensure adhesion and cross-linking. The pseudo logic of the pathing can be summarized as: in the pathing code, letting $R$ be interconnected region and a boundary $\partial R$. Establish a Euclidean distance transformation $D_R$:

$$D_R(p) = \min_{q \in \partial R} ||p - q|| \quad \forall p \in R$$

Iso-contours $C_d$ are level sets of $D_R$ that are defined as:

$$C_d = \{p \in R \mid D_R(p) = d\}$$

Routing and rerouting are performed between adjacent iso-contours $C_{d_1}$ and $C_{d_2}$, where ($d_1 < d_2$), by connecting to form a spiral and by defining rerouting points. For a point $p \in C_{d_1}$, proceed to find its inward link $I(p)$:

$$I(p) = \arg \min_{q \in C_{d_2}} ||p - q||$$

As well as the outward link $O(p)$:

$$O(p) = \arg \min_{q \in C_{d_0}} ||p - q||, d_0 < d_1$$

Beginning at point $p_0 \in \partial R$ connect points recursively with:

$$p_{i+1} = \begin{cases} I(p_i), & \text{if moving inward} \\ O(p_i), & \text{if moving outward after center is reached} \end{cases}$$

Form the Fermat spiral array $\pi$ that contains both inward $I(p)$ and outward $O(p)$ traversal. Where $r(t)$ and $\theta(t)$ are examples of angular coordinates assigned by the rerouting process.

$$\pi(t) = \begin{cases} r(t)\cos(\theta(t)), & \text{inward path} \\ r(t)\sin(\theta(t)), & \text{outward path} \end{cases}$$

Unlike the fully implemented and smoothed connected Fermat pathing protocol, our raw pathed points are interpolated using a B-spline fit through the Scipy python package *splev*.

## Acknowledgments

We gratefully acknowledge support from the NSF through the Harvard MRSE (DMR-2011754) and the ARO MURI program (W911NF-22-1-0219). We also thank Giada Risso, Leon Kamp, and Prof. Katia Bertoldi for their insightful discussions on the bilayer structure design. We would also like to acknowledge Dr. Haisen Zhao for providing the published and open format of the Fermat Pathing toolkit software and Dr. Andrew Spielberg for assisting in compiling the connected Fermat pathing toolkit for use on local machines. JKW is supported by the National Science Foundation Graduate Research Fellowship.

**Data availability:**

[Google collab Python code database](#)

**Competing interests:**

Natalie M. Larson, Jochen Mueller, Jennifer A. Lewis, "Printhead and method of printing multimaterial filaments including oriented, twisted and/or helical features," Patent No.: US 12,005,631 B2, Date of Patent: June 11, 2024. (Harvard University ref. 8294)

REFERENCES

[1]  K. Suzumori, S. Iikura, H. Tanaka, *Proceedings. 1991 IEEE International Conference on Robotics and Automation* **1991**, 1622.
[2]  R. F. Shepherd, F. Ilievski, W. Choi, S. A. Morin, A. A. Stokes, A. D. Mazzeo, X. Chen, M. Wang, G. M. Whitesides, *Proc. Natl. Acad. Sci. U.S.A.* **2011**, *108*, 20400.
[3]  F. Ilievski, A. D. Mazzeo, R. F. Shepherd, X. Chen, G. M. Whitesides, *Angew. Chem.* **2011**, *123*, 1930.
[4]  R. V. Martinez, J. L. Branch, C. R. Fish, L. Jin, R. F. Shepherd, R. M. D. Nunes, Z. Suo, G. M. Whitesides, *Adv. Mater.* **2013**, *25*, 205.
[5]  B. Mosadegh, P. Polygerinos, C. Keplinger, S. Wennstedt, R. F. Shepherd, U. Gupta, J. Shim, K. Bertoldi, C. J. Walsh, G. M. Whitesides, *Adv. Funct. Mater.* **2014**, *24*, 2163.
[6]  D. Rus, M. T. Tolley, *Nature* **2015**, *521*, 467.
[7]  C. Majidi, *Adv Materials Technologies* **2019**, *4*, 1800477.
[8]  E. W. Hawkes, C. Majidi, M. T. Tolley, *Sci. Robot.* **2021**, *6*, eabg6049.
[9]  E. Siéfert, E. Reyssat, J. Bico, B. Roman, *Nature Mater* **2019**, *18*, 24.
[10] T. J. Jones, E. Jambon-Puillet, J. Marthelot, P.-T. Brun, *Nature* **2021**, *599*, 229.
[11] T. J. Wallin, J. Pikul, R. F. Shepherd, *Nat Rev Mater* **2018**, *3*, 84.
[12] R. L. Truby, J. A. Lewis, *Nature* **2016**, *540*, 371.
[13] M. A. Skylar-Scott, J. Mueller, C. W. Visser, J. A. Lewis, *Nature* **2019**, *575*, 330.
[14] M. Wehner, R. L. Truby, D. J. Fitzgerald, B. Mosadegh, G. M. Whitesides, J. A. Lewis, R. J. Wood, *Nature* **2016**, *536*, 451.
[15] N. M. Larson, J. Mueller, A. Chortos, Z. S. Davidson, D. R. Clarke, J. A. Lewis, *Nature* **2023**, DOI 10.1038/s41586-022-05490-7.
[16] N. M. Larson, *MRS Bulletin* **2024**, *49*.
[17] L. Ren, W. Li, H. Liu, B. Li, X. Zhou, L. Ren, Z. Han, Z. Song, Q. Liu, *Additive Manufacturing* **2023**, *73*, 103661.
[18] T. Pleij, A. V. Bayles, J. Vermant, *Adv Materials Technologies* **2024**, *9*, 2400005.
[19] W. Wu, A. DeConinck, J. A. Lewis, *Advanced Materials* **2011**, *23*, DOI 10.1002/adma.201004625.
[20] H. Zhao, F. Gu, Q.-X. Huang, J. Garcia, Y. Chen, C. Tu, B. Benes, H. Zhang, D. Cohen-Or, B. Chen, *ACM Trans. Graph.* **2016**, *35*, 1.

**Supporting Information**

**Rotational Multi-material 3D Printing of Soft Robotic Matter with Asymmetrical Embedded Pneumatics**

**Jackson K. Wilt[1*], Natalie M. Larson[2*], and Jennifer A. Lewis[1,3] †**

[1]John A. Paulson School of Engineering and Applied Sciences, Harvard University, Cambridge, MA 02138, USA.

[2]Department of Mechanical Engineering, Stanford University, Stanford, CA 94305, USA.

[3]Wyss Institute for Biologically Inspired Engineering, Harvard University, Cambridge, MA 02138, USA.

*Authors contributed equally to this work

† Corresponding author

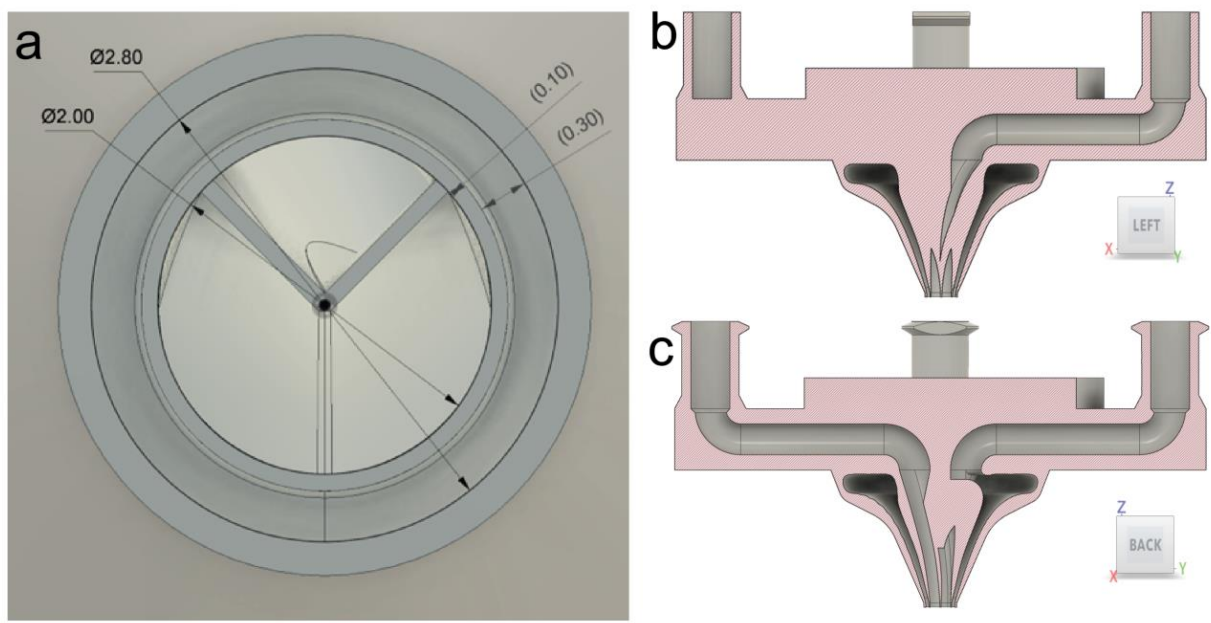

**Figure S1**: Nozzle geometry. (a) Cross section of printed nozzle geometry containing the three channels and internal dimensions. (b) "Left" side section cut from the nozzle and (c) "back" side section cut of the nozzle.

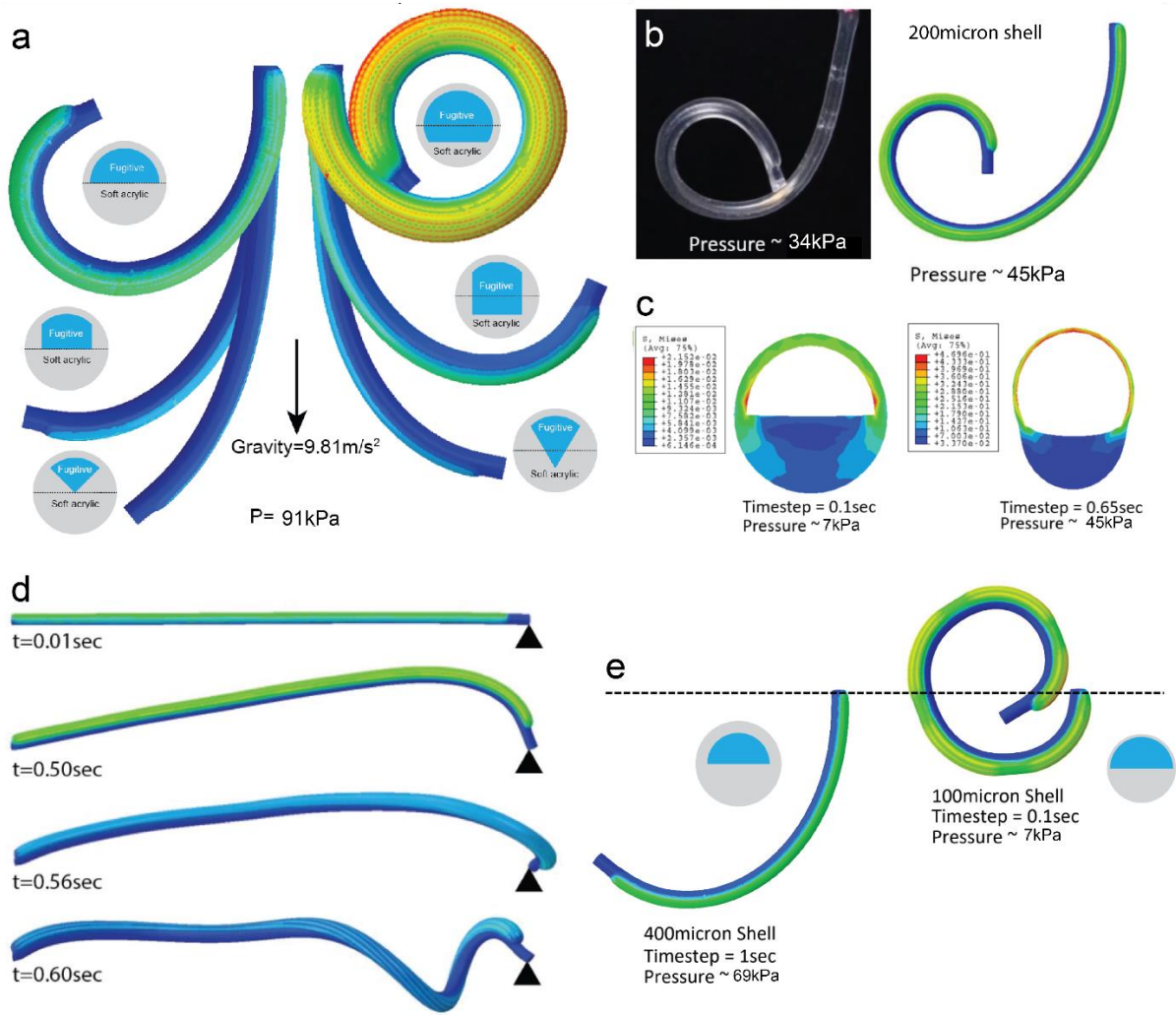

**Figure S2**: Actuation properties of varying cross-sectional geometry. (a) An array of different cross-sectional pressure responses of 50mm rod using Abaqus. (b) Actuated $\varphi = 0º$ and 50mm filament actuated with air pressure and visualized with Abaqus. (c) Internal geometry is visualized in the early stages of pressurization (left) and fully pressurized cross-section. (d) Observation of nonlinear and out-of-plane deflection of a long pinned actuated filament. (e) Actuating the $\varphi = 0º$ with different shell thicknesses 400μm to 100μm.

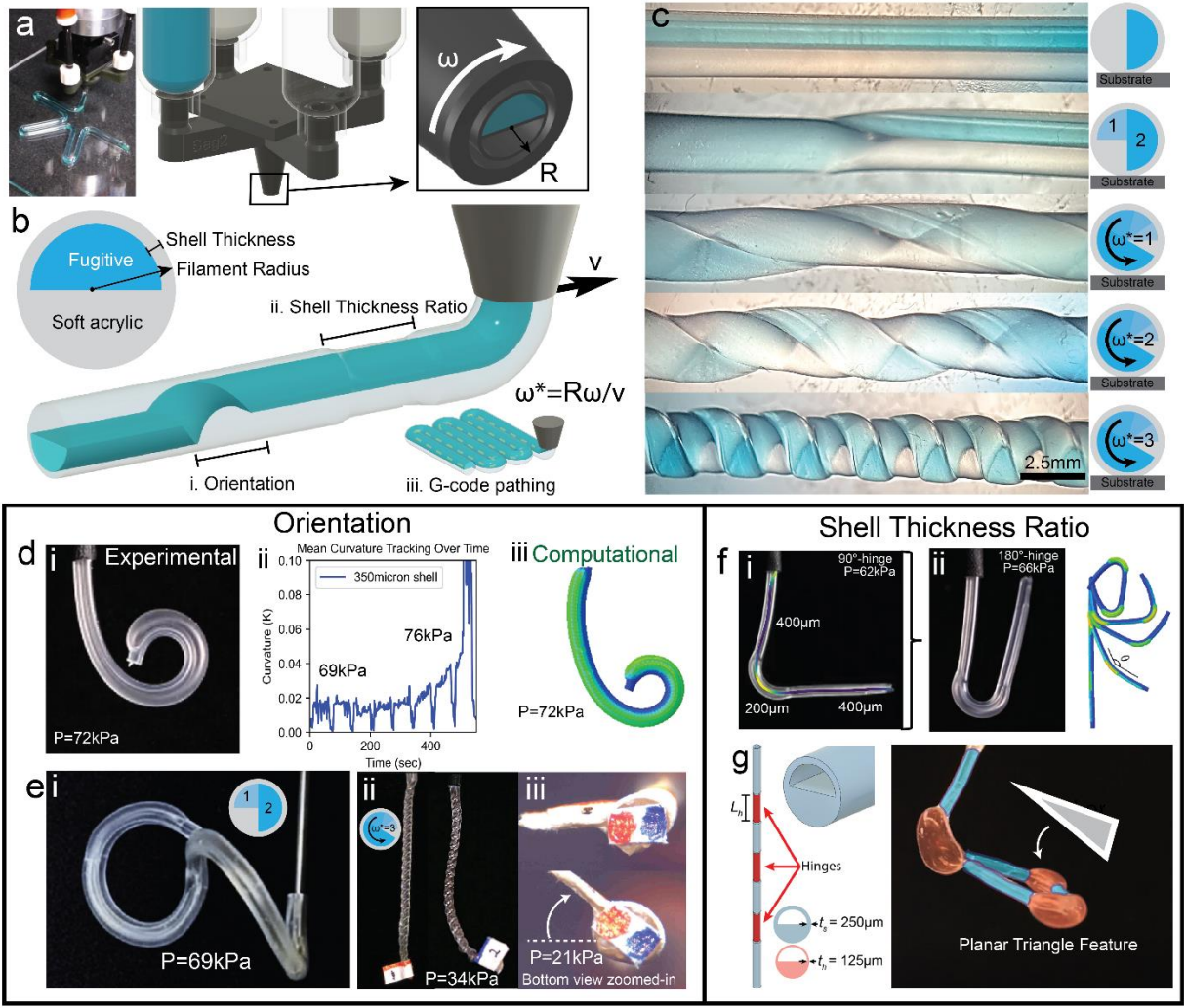

**Figure S3**: Additional properties of $\varphi = 0º$ nozzle geometry. (a) Printing process and (b) diagram of rotational printing with correlating print parameters: orientation and shell thickness ratio in combination with the print pathing. (c) Printed and (d) actuated reoriented filaments include discrete and continuous reorientation characteristics. (d,i) Experimental and (d,ii) curvature response to increasing actuation pressure to (d,iii) computational responses of a 50mm long actuator with progressively increasing pressure. (e,i) Actuation of the discrete oriented filament by 90º reorientation to (e,ii) continuously oriented filaments along with the (e,iii) bottom end imaged. (h) Modulation of elastomer shell thickness that produces (f,i) 90º and (f,ii) 180º hinge angle. (g) A computational diagram of regions with changing shell thickness and plane of actuation that produces multihinged actuators both experimentally.

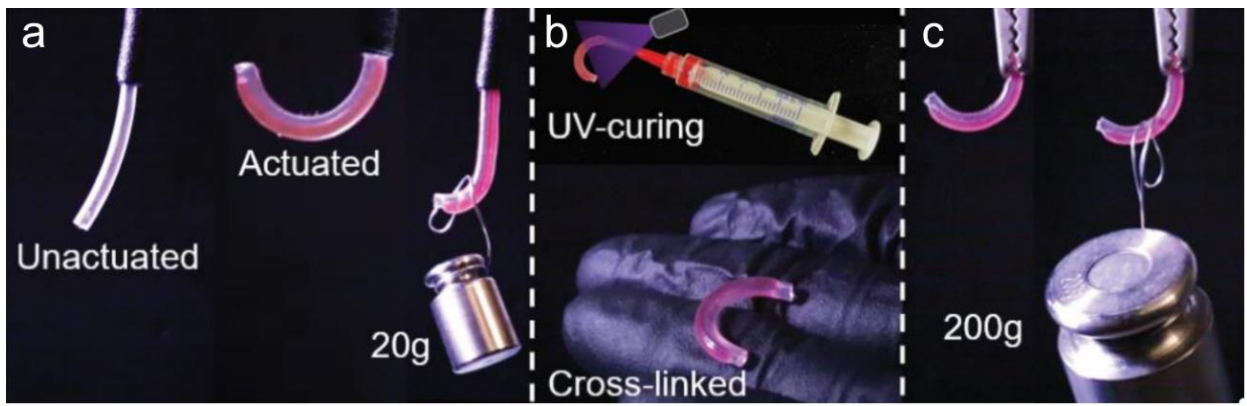

**Figure S4**: Injection and actuation of rigidly curing resin material. (a) Actuation using liquid rigidly curing resin holding 20g weight. (b) An image of a syringe is used for inflation and used for the fully cross-linked state. (c) The crosslinked actuated filament holds a 200g weight

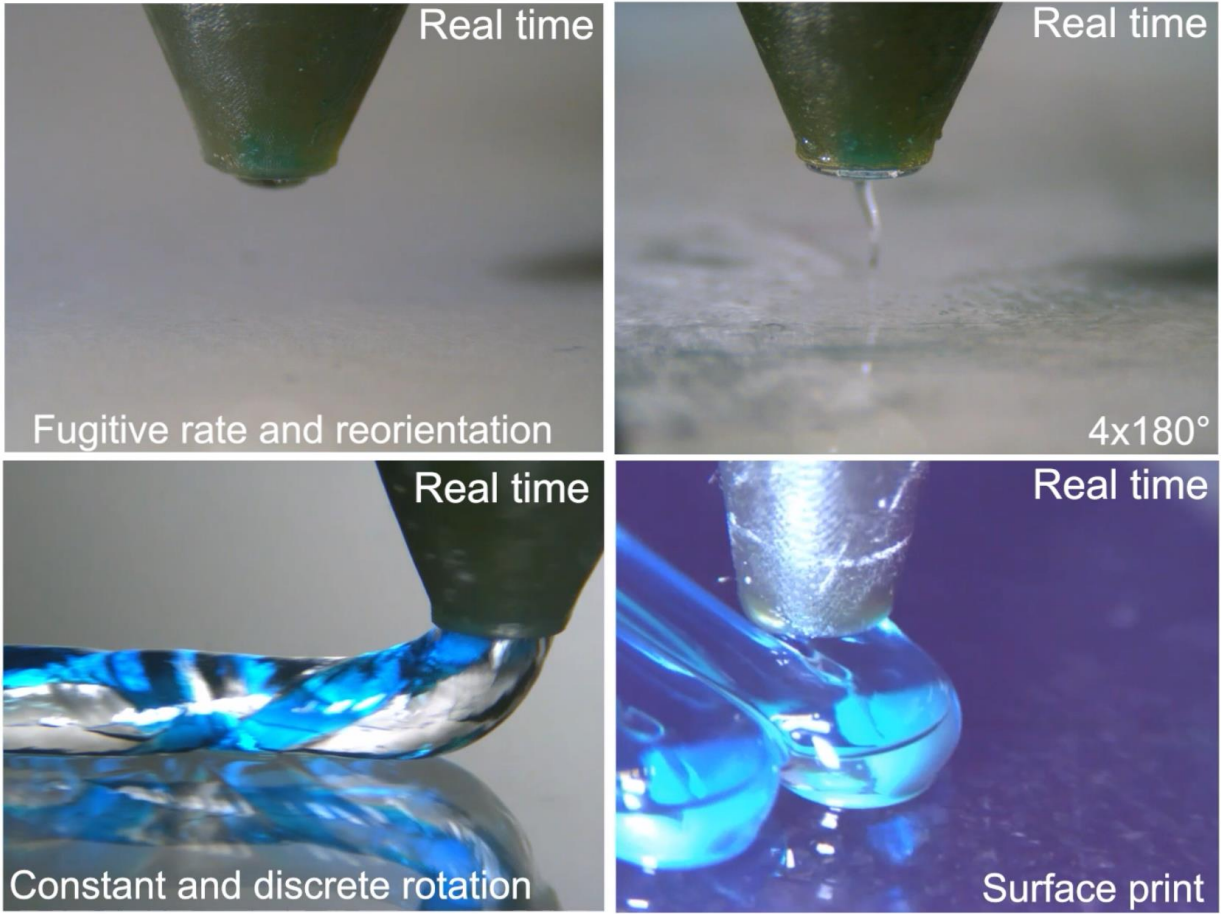

**Movie 1**: Example of printing the filaments and surfaces in real-time.

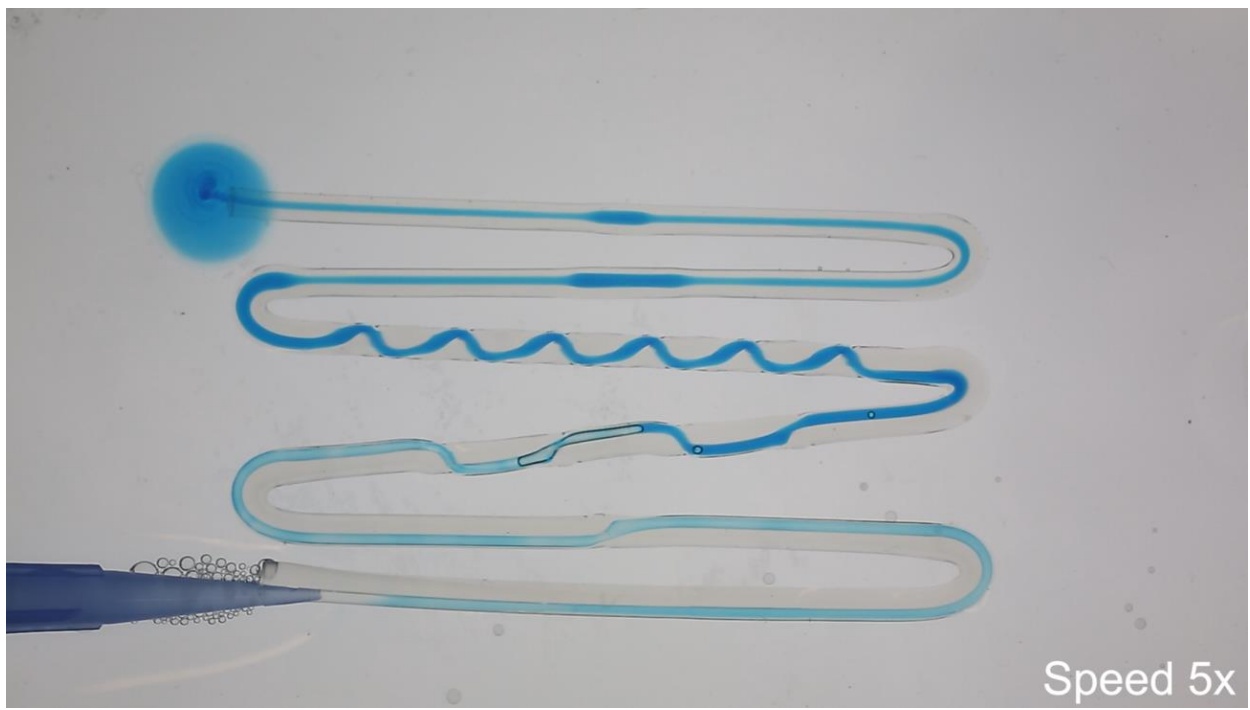

**Movie 2**: Perfusion procedure of a continually extruded filament placed in an ice water bath.

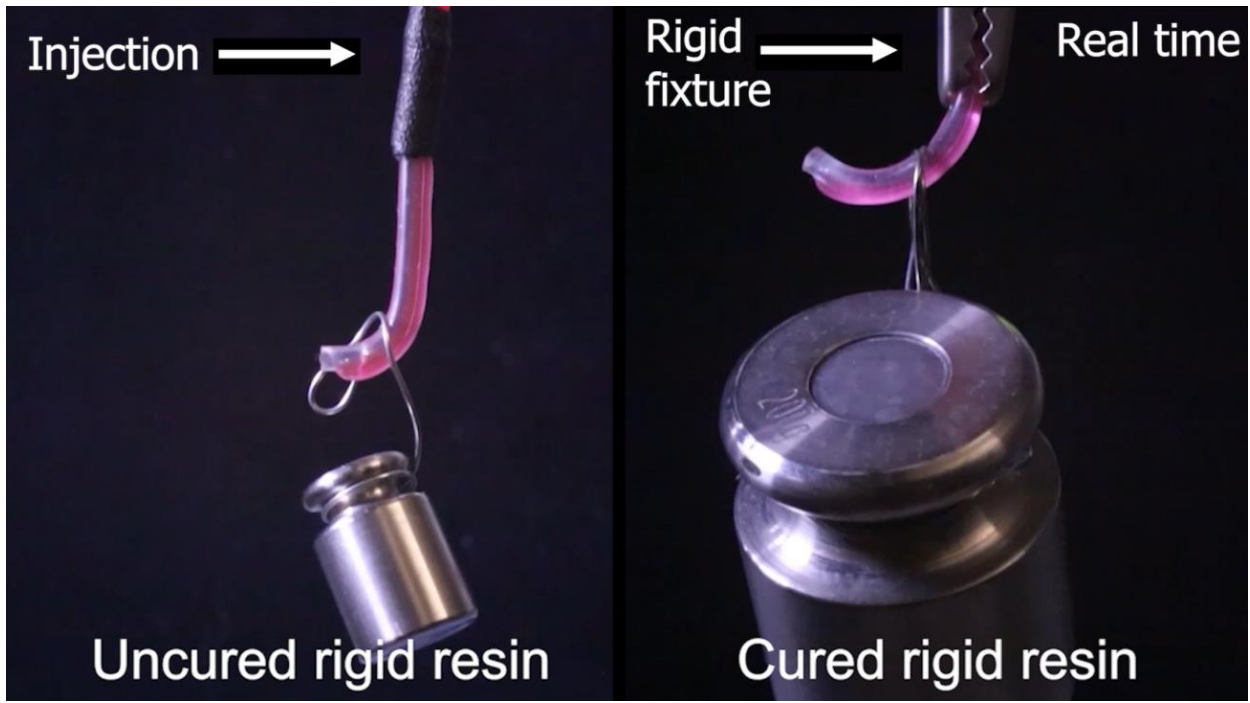

**Movie 3**: Injection and actuation of rigidly curing resin material.

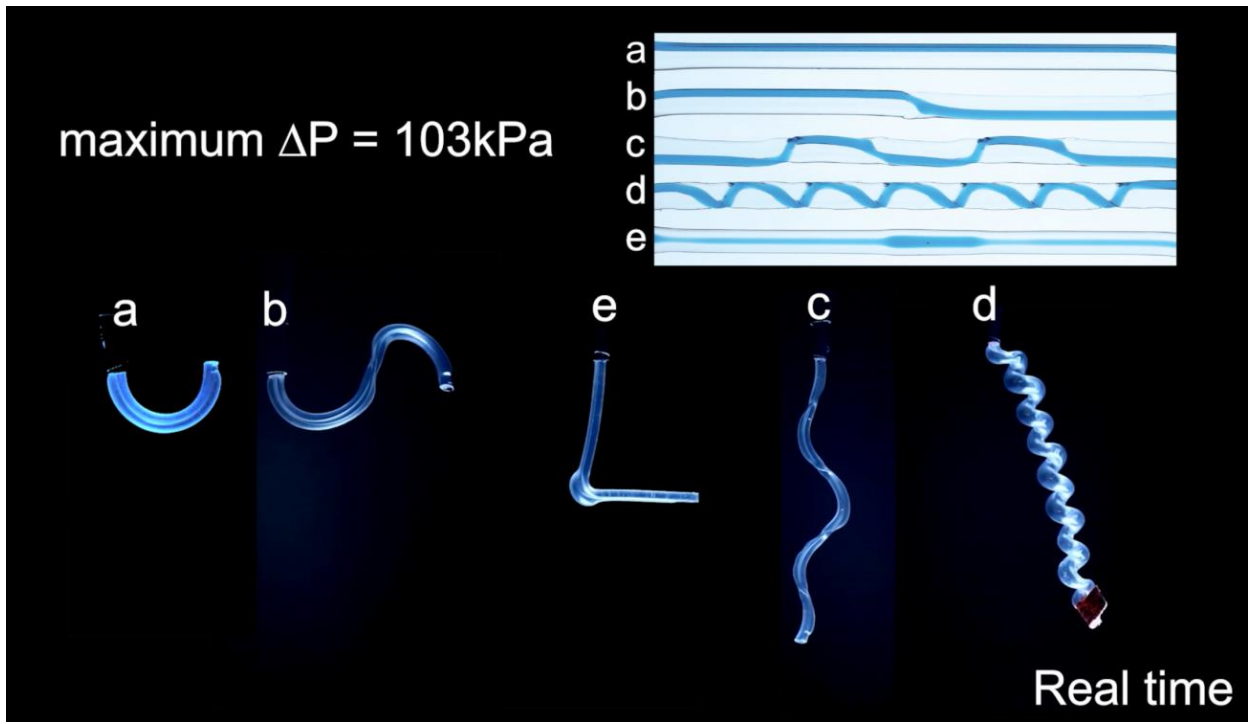

**Movie 4**: Series of actuating 1D filaments being volumetrically and pneumatically inflated.

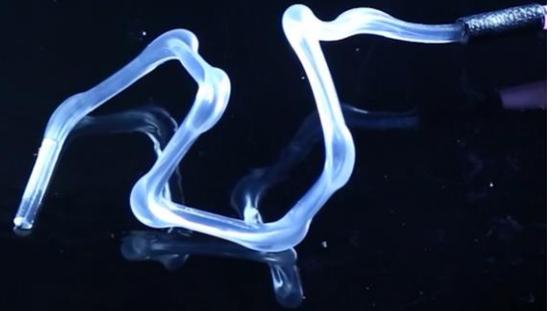

**Movie 5**: Wireframe cube actuator being pressurized.

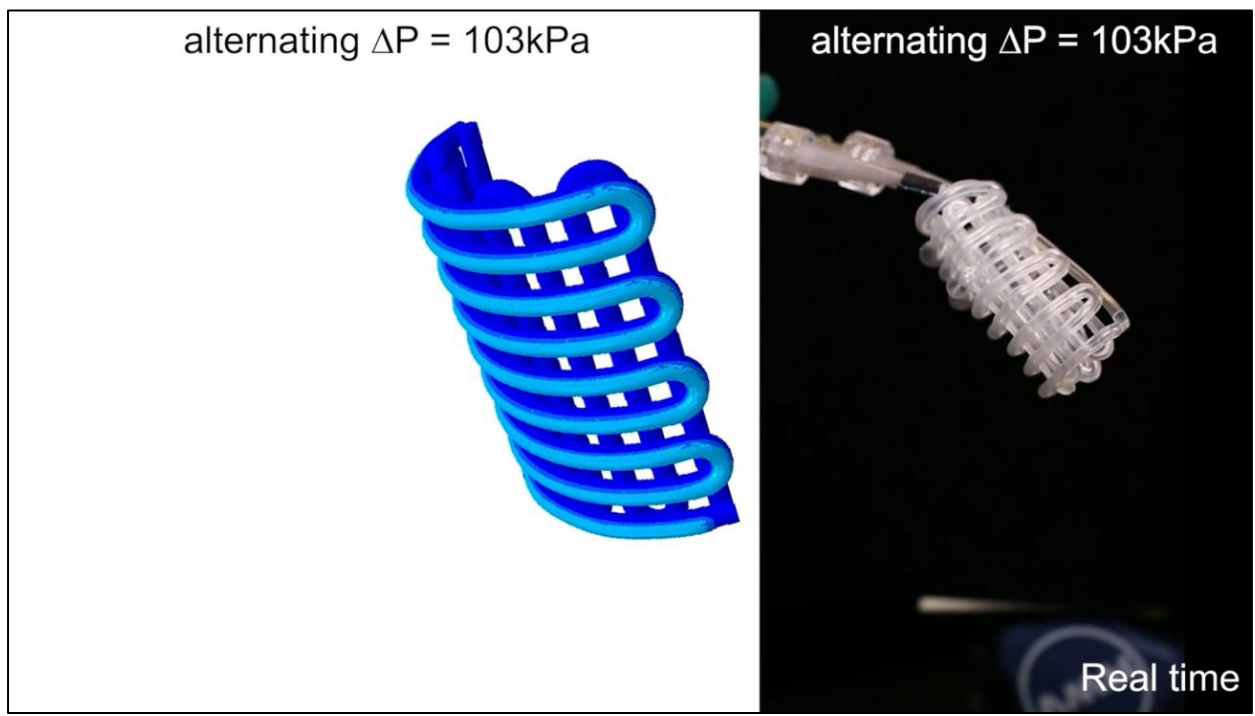

**Movie 6**: Bilayer competing actuation in FEA and experimental results.

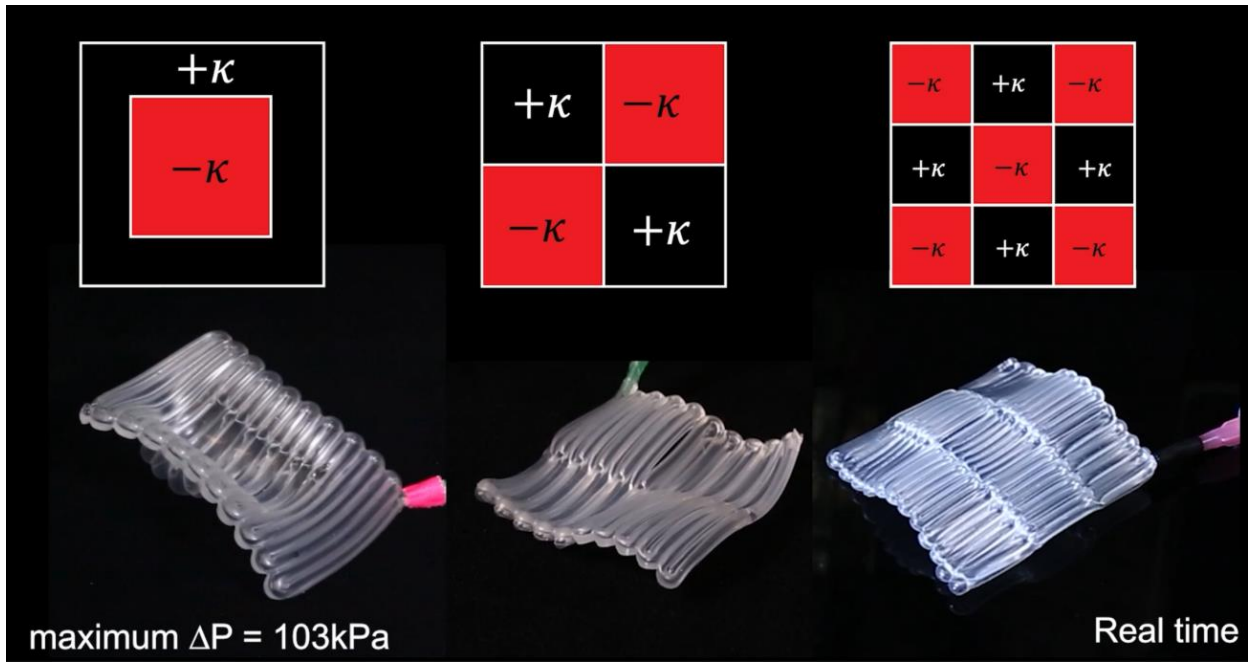

**Movie 7**: Surface actuating video for various rectilinear regional assignments of discrete curvatures.

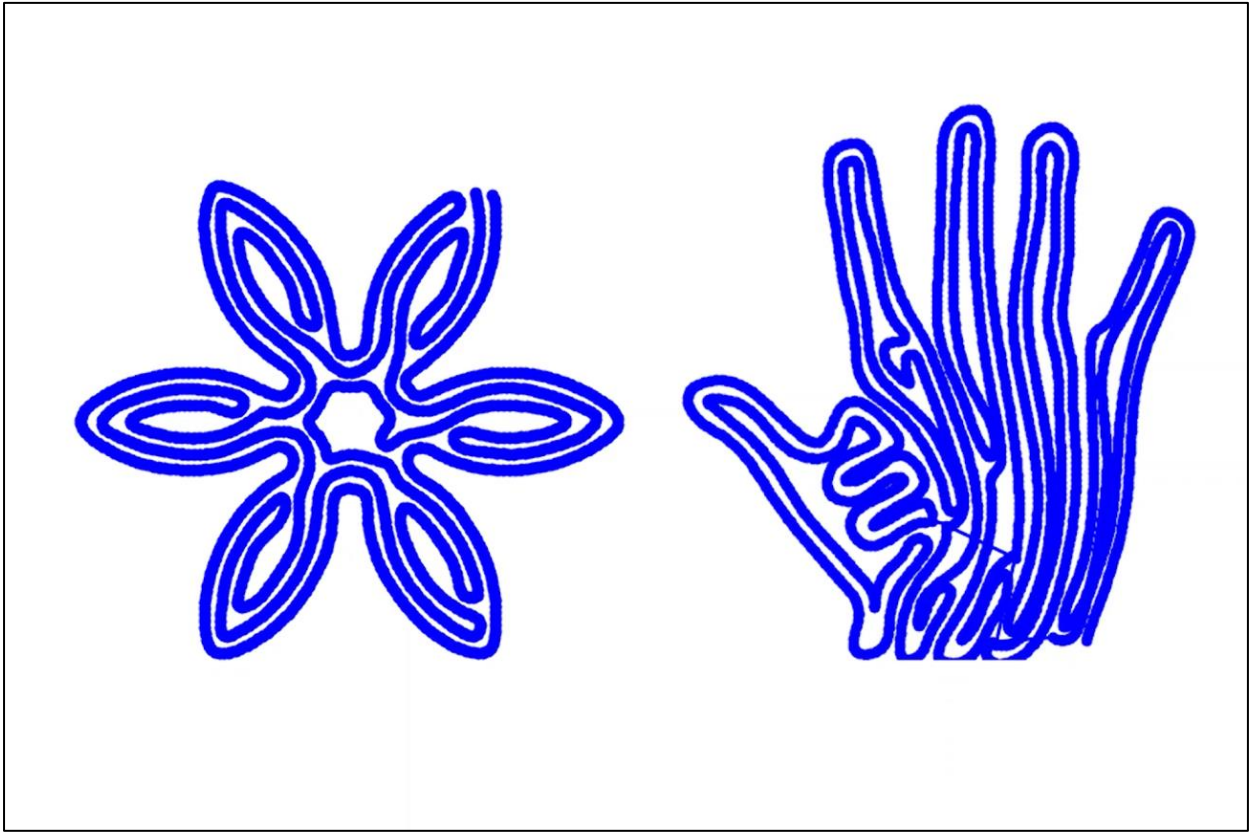

**Movie 8**: Connected Fermat Spiral code-based pathing the contours of the exemplar flower and hand designs.

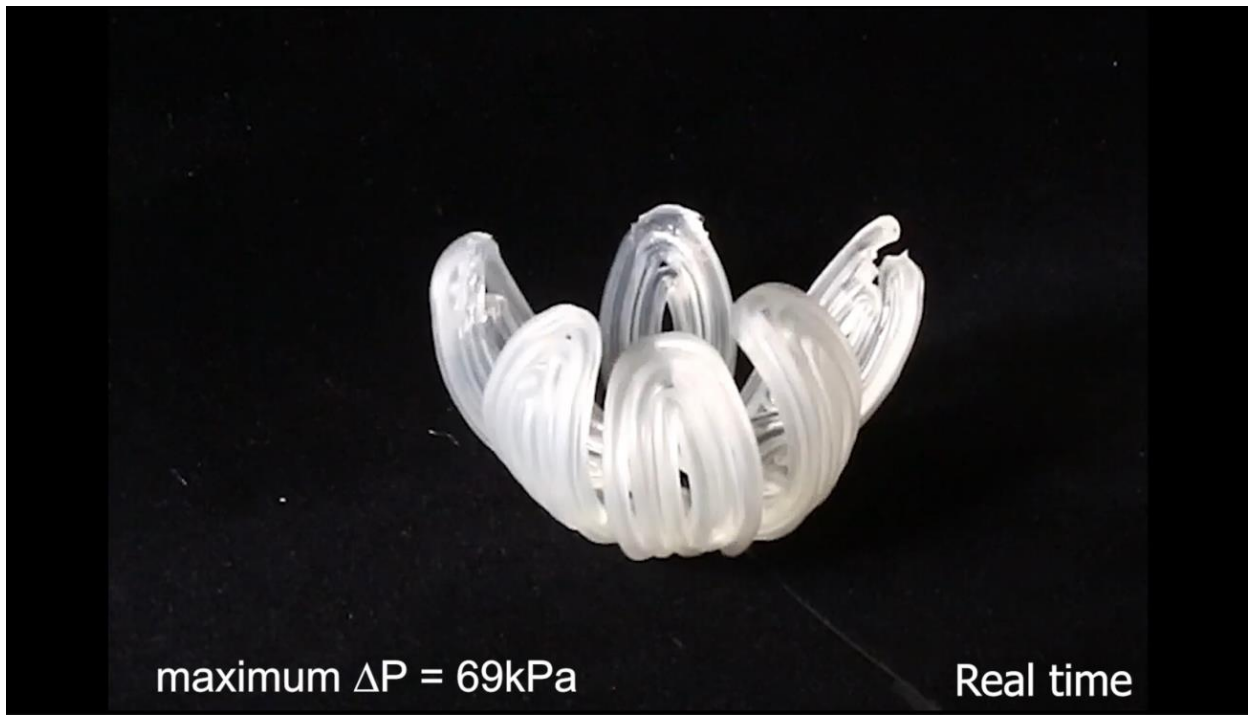

**Movie 9**: Compilation video of the full printing process using the connected Fermat spiral pathing with a flower-like vectorized image to produce a printing to actuator workflow.

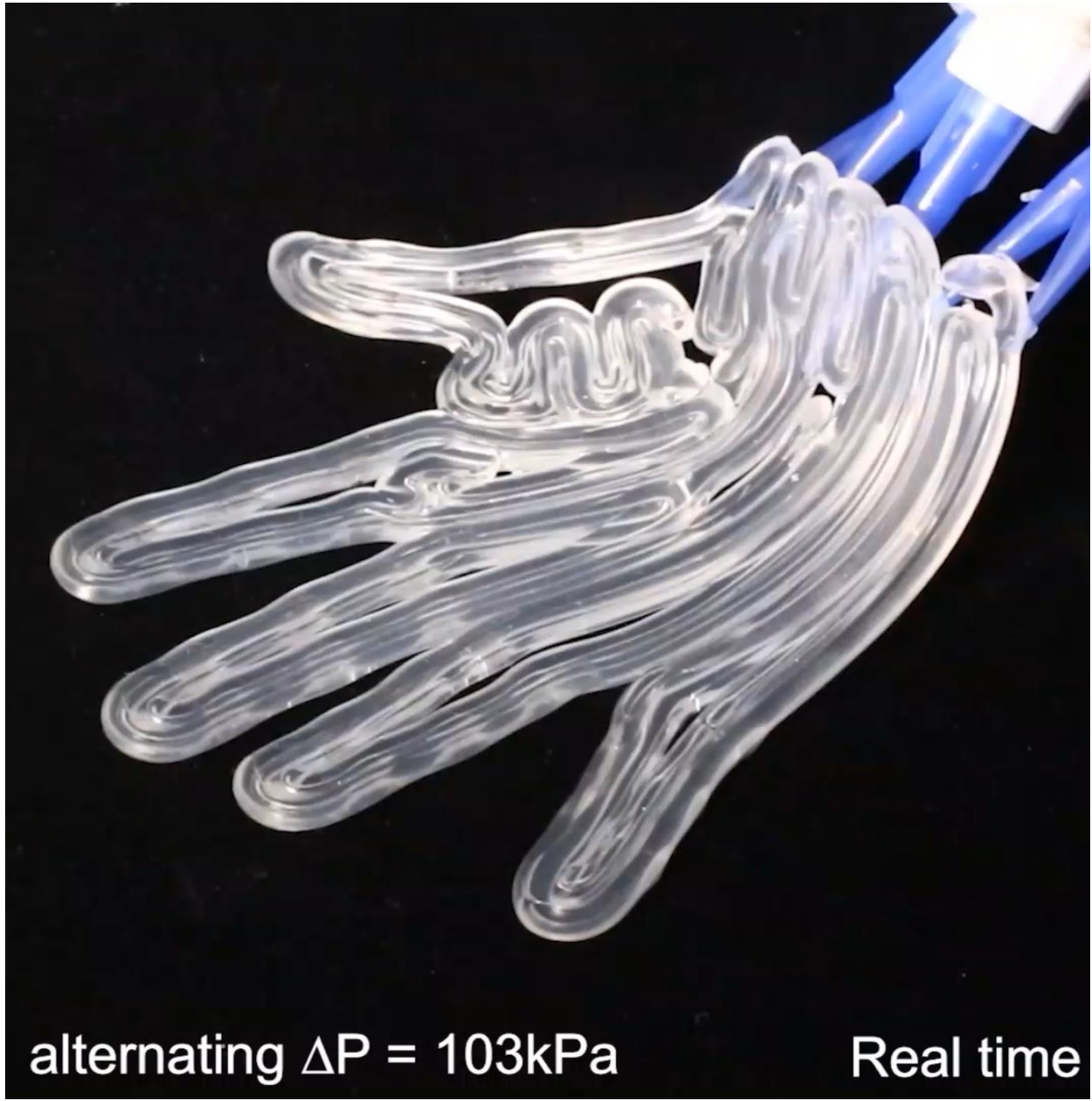

**Movie 10**: Hand actuation of each digit using pressurized air input in sequence: pinky, ring, middle, index, and thumb finger.